\documentclass{article}
\usepackage{amsmath}

\input{tcilatex}

\begin{document}

\begin{center}
{\huge Integrable systems of quartic oscillators. II}

\bigskip

\textbf{M. Bruschi}$^{\ast }$\ and \textbf{F. Calogero}$^{+}$

Dipartimento di Fisica, Universit\`{a} di Roma "La Sapienza", 00185 Roma,
Italy

Istituto Nazionale di Fisica Nucleare, Sezione di Roma

\bigskip

\textit{Abstract}
\end{center}

Several completely \textit{integrable}, indeed \textit{solvable},
Hamiltonian many-body problems are exhibited, characterized by Newtonian
equations of motion (\textquotedblleft acceleration equal
force\textquotedblright ), with linear and cubic forces, in $N$-dimensional
space ($N$ being an arbitrary positive integer, with special attention to $%
N=2$, namely motions in a plane, and $N=3$, namely motions in ordinary
three-dimensional space). All the equations of motion are written in
covariant form (\textquotedblleft $N$-vector equal $N$-vector%
\textquotedblright ), entailing their rotational invariance. The
corresponding Hamiltonians are of normal type, with the kinetic energy
quadratic in the canonical momenta, and the potential energy quadratic and
quartic in the canonical coordinates.

\bigskip


Physics and Astronomy Classification Scheme: 05.45.-a, 45.05+x, 45.50.Jf

\bigskip

\bigskip\ 

\bigskip

\bigskip

\bigskip

\bigskip

$^{+}$Corresponding author. Fax: +39-06-4957697

$^{\ast }$mario.bruschi@roma1.infn.it

$^{+}$francesco.calogero@roma1.infn.it, francesco.calogero@uniroma1.it\qquad

\newpage

\section{Introduction}

In a recent paper \cite{1} we introduced several \textit{integrable}, indeed 
\textit{solvable}, nonlinear oscillators. The basic idea was to start from
the \textit{integrable}, indeed \textit{solvable}, matrix ODE \cite{2} \cite%
{3} \cite{4}~\cite{K}%
\begin{equation}
\underline{\ddot{U}}=-\left( \underline{A}\,\underline{U}+\underline{U}\,%
\underline{A}+\underline{U}^{3}\right) ~,  \label{1}
\end{equation}%
and to transform it into a system of \textit{covariant}, hence \textit{%
rotation-invariant}, Newtonian equations ("acceleration equal force", with
linear and cubic forces) for the time evolution of $N$-vectors, by
parameterizing appropriately the matrix $\underline{U}$\ in terms of these $%
N $-vectors. In the present paper we follow the same approach \cite{1} \cite%
{3} \cite{4}, but we exploit certain additional, convenient
parameterizations of matrices in terms of vectors, which we presented
recently \cite{5} and which also transform the matrix evolution equation (%
\ref{1}) into \textit{covariant}, hence \textit{rotation-invariant},
Newtonian equations of motions for $N$-vectors in $N$-dimensional space. We
thereby obtain -- and exhibit in the following Section 2 -- the Newtonian
equations of motion of several new \textit{integrable}, indeed \textit{%
solvable}, systems of oscillators subjected to linear and cubic forces. As
shown in Section 3, all these systems are Hamiltonian, with standard
Hamiltonians of normal form: the sum of a standard kinetic energy quadratic
in the momenta and a potential energy depending (quadratically and
quartically) only on the canonical coordinates. In view of the
phenomenological relevance of such nonlinear oscillators, it is remarkable
that several instances of such systems exist which are \textit{integrable}
indeed \textit{solvable}. Investigation of their detailed dynamics is
postponed to future papers.

\emph{Notation}: throughout this paper matrices are denoted by underlined
characters, and vectors by superimposed arrows; their dimensionality will in
each case be clear from the context. Superimposed dots indicate
differentiations with respect to the independent variable ("time"), while
dots sandwiched among two vectors denote the standard scalar product. As a
rule, we write the equations of motion in the neatest form, forsaking any
attempt to introduce additional constants by rescaling the independent
variable and/or by redefining the dependent variables via linear
transformations.

\bigskip

\section{Newtonian equations of motion}

A convenient way to obtain \textit{covariant} ("vector equal vector") hence 
\textit{rotation-invariant} Newtonian equations of motion is to insert in
the matrix evolution equation (\ref{1}) the parameterizations introduced in 
\cite{5} (hereafter the equation (n) of this paper \cite{5} is identified
here as (I.n)). Since the parameterization P1 (I.3) was already exploited in 
\cite{1}, we exhibit here only the new dynamical systems yielded by the new
parameterization P2 (I.4). Let us emphasize that, since the matrix evolution
equation (\ref{1}) is \textit{integrable}, indeed \textit{solvable}, all the
dynamical systems obtained in this manner are as well \textit{integrable},
indeed \textit{solvable}.

Let us begin by writing down our basic result, obtained by inserting the\
formulas (I.4) into (\ref{1}) and by then performing some appropriate,
merely notational, changes: 
\begin{subequations}
\label{2}
\begin{equation}
\overset{..}{\overrightarrow{r}}^{(n\ell )}=-\sum_{\nu =1}^{N}\sum_{\lambda
=1}^{L}\left\{ \overrightarrow{r}^{(\nu \lambda )}\,\left[ a_{\lambda \ell
}^{(\nu n)}+\left( \overrightarrow{\tilde{r}}^{(\nu \lambda )}\cdot 
\overrightarrow{r}^{(n\ell )}\right) \right] \right\} ~,  \label{2a}
\end{equation}

\begin{equation}
\overset{..}{\overrightarrow{\tilde{r}}}^{(n\ell )}=-\sum_{\nu
=1}^{N}\sum_{\lambda =1}^{L}\left\{ \overrightarrow{\tilde{r}}^{(\nu \lambda
)}\,\left[ a_{\ell \lambda }^{(n\nu )}+\left( \overrightarrow{r}^{(\nu
\lambda )}\cdot \overrightarrow{\tilde{r}}^{(n\ell )}\right) \right]
\right\} ~.  \label{2b}
\end{equation}%
Here, and throughout, the index $\ n$ (as well of course as $\nu )$ ranges
from $1$ to $N,$ and$\,$ the index $\ell $ (as well of course as $\lambda )$
ranges from $1$ to $L,$ with $N$ and $L$ two arbitrary positive integers;
the $2\,N\,L$ time-dependent quantities $\overrightarrow{r}^{(n\ell )}\equiv 
\overrightarrow{r}^{(n\ell )}(t),$ $\overrightarrow{\tilde{r}}^{(n\ell
)}\equiv \overrightarrow{\tilde{r}}^{(n\ell )}(t)$ are $N$-vectors
(identifying the positions of moving point particles in $N$-dimensional
space); the dots sandwiched among $N-$vectors denote the standard scalar
product; and the $(N\,L)^{2}$ quantities $a_{\ell \lambda }^{(n\nu )}$ are
arbitrary scalar constants. Because of the way they have been obtained,
these $2\,N\,L$ Newtonian equations of motion, featuring linear and cubic
forces, are \textit{integrable} indeed \textit{solvable}; and they are
clearly \textit{covariant} ($N$-vector equal $N$-vector), hence \textit{%
rotation-invariant }in $N$-dimensional space.

It would seem that this system always involves a number of $N$-vectors which
is an even multiple of the number of dimensions $N$ of the vector space. But
this is not the case, due to the possibility of reductions, which may be
enforced (with an appropriate choice of initial conditions) provided the
constants $a_{\ell \lambda }^{(n\nu )}$ satisfy appropriate restrictions.
For instance if these constants satisfy the symmetry restriction 
\end{subequations}
\begin{subequations}
\label{3}
\begin{equation}
a_{\ell \lambda }^{(n\nu )}=a_{\lambda \ell }^{(\nu n)}~,  \label{3a}
\end{equation}%
it is clearly possible to impose the corresponding reduction%
\begin{equation}
\overrightarrow{\tilde{r}}^{(n\ell )}=\overrightarrow{r}^{(n\ell )}~,
\label{3b}
\end{equation}%
which obviously halves (from $2\,N\,L$ to $N\,L)$ the number of evolving
vectors, as well as the number of $N$-vector Newtonian equations of motion,
which clearly read then (see (\ref{2}))%
\begin{equation}
\overset{..}{\overrightarrow{r}}^{(n\ell )}=-\sum_{\nu =1}^{N}\sum_{\lambda
=1}^{L}\left\{ \overrightarrow{r}^{(\nu \lambda )}\,\left[ a_{\lambda \ell
}^{(\nu n)}+\left( \overrightarrow{r}^{(\nu \lambda )}\cdot \overrightarrow{r%
}^{(n\ell )}\right) \right] \right\} ~.  \label{3c}
\end{equation}%
Other (or additional) reductions of the Newtonian equations of motion (\ref%
{2}) are moreover possible, in which some of the $N$-vectors vanish
identically, since clearly, provided for some value of the indices $n$ and $%
\ell $ the constants $a_{\lambda \ell }^{(\nu n)}$ vanish unless $\nu =n$
and $\lambda =\ell $ ($a_{\lambda \ell }^{(\nu n)}=\delta _{n\nu }\,\delta
_{\ell \lambda }\,a_{\ell }^{(n)}~$for $\nu =1,...,N;~\lambda =1,...,L~$with 
$n,\ell $ fixed), then it is consistent, see (\ref{2a}), to set $%
\overrightarrow{r}^{(n\ell )}=0;$ and likewise, provided for some value of
the indices $n$ and $\ell $ the constants $a_{\ell \lambda }^{(n\nu )}$
vanish unless $\nu =n$ and $\lambda =\ell $ ($a_{\ell \lambda }^{(n\nu
)}=\delta _{n\nu }\,\delta _{\ell \lambda }\,a_{\ell }^{(n)}~$for $\nu
=1,...,N;~\lambda =1,...,L~$with $n,\ell $ fixed), then it is consistent,
see (\ref{2b}), to set $\overrightarrow{\tilde{r}}^{(n\ell )}=0.$

Let us also point out that, if all the constants $a_{\ell \lambda }^{(n\nu
)} $ vanish, so that the Newtonian equations of motion (\ref{2}) only
involve cubic forces, by applying an argument analogous to that used in \cite%
{6} one concludes that \textit{all the nonsingular solutions of the
following (}$\omega $\textit{-deformed, complex)} \textit{Newtonian
equations of motion,} 
\end{subequations}
\begin{subequations}
\label{4}
\begin{equation}
\overset{..}{\overrightarrow{r}}^{(n\ell )}-3\,i\,\omega \,\overset{.}{%
\overrightarrow{r}}^{(n\ell )}-2\,\omega ^{2}\,\overrightarrow{r}^{(n\ell
)}=c\,\sum_{\nu =1}^{N}\sum_{\lambda =1}^{L}\left[ \overrightarrow{r}^{(\nu
\lambda )}\,\left( \overrightarrow{\tilde{r}}^{(\nu \lambda )}\cdot 
\overrightarrow{r}^{(n\ell )}\right) \right] ~,  \label{4a}
\end{equation}

\begin{equation}
\overset{..}{\overrightarrow{\tilde{r}}}^{(n\ell )}-3\,i\,\omega \,\overset{.%
}{\overrightarrow{\tilde{r}}}^{(n\ell )}-2\,\omega ^{2}\,\overrightarrow{%
\tilde{r}}^{(n\ell )}=c\,\sum_{\nu =1}^{N}\sum_{\lambda =1}^{L}\left[ 
\overrightarrow{\tilde{r}}^{(\nu \lambda )}\,\left( \overrightarrow{r}^{(\nu
\lambda )}\cdot \overrightarrow{\tilde{r}}^{(n\ell )}\right) \right] ~.
\label{4b}
\end{equation}%
\textit{are completely periodic with period }$2\,\pi \,/\,\omega $\textit{,
provided }$\omega $\textit{\ is a nonvanishing real (without loss of
generality, positive) constant}. Note that these equations of motion are 
\textit{complex} (see the second term in their left-hand sides); but \textit{%
real} equations can be obtained from these by introducing the real and
imaginary parts of the $N$-vectors $\overrightarrow{r}^{(n\ell )}$ and $%
\overrightarrow{\tilde{r}}^{(n\ell )},$ and of the "coupling constant" $c$
(which has been introduced in (\ref{4}) via a trivial rescaling of the
dependent variables), and by then considering separately the real and
imaginary parts of these equations. There thus results a doubling (from $%
2\,N\,L$ to $4\,N\,L$) of the number of evolving \textit{real }$N$-vectors;
but this number can be cut down via the same kind of reductions discussed
above. No additional elaboration is reported below on these systems of 
\textit{nonlinear harmonic oscillators }\cite{6}.

In the rest of this Section 2 we display explicitly, in the physically more
interesting cases with $N=2$ (evolutions in the plane) and $N=3$ (evolutions
in ordinary, three-dimensional, space), some simple examples involving few
moving points. For the sake of completeness, and for the reader's
enlightenment, we also display, in self-evident notation \cite{5}, the
parameterizations that relate the equations of motion we report to the basic
matrix evolution equation (\ref{1}).

\bigskip

\subsection{Oscillators in the plane}

Let us consider the following parameterization of a $4\otimes 4$ matrix $%
\underline{U}$ (see (\ref{1})) in terms of \textit{four} $2-$vectors (see
(I.6)): 
\end{subequations}
\begin{subequations}
\label{5}
\begin{equation}
\underline{U}\doteq \left( \overrightarrow{r}^{\left( 1\right) },%
\overrightarrow{r}^{\left( 2\right) };\overrightarrow{\tilde{r}}^{\left(
1\right) },\overrightarrow{\tilde{r}}^{\left( 2\right) }\right) ~,
\label{5a}
\end{equation}%
\begin{equation}
\overrightarrow{r}^{\left( n\right) }\equiv \left( x^{\left( n\right)
},y^{\left( n\right) }\right) ,~~~\overrightarrow{\tilde{r}}^{\left(
n\right) }\equiv \left( \tilde{x}^{\left( n\right) },\tilde{y}^{\left(
n\right) }\right) ,~~~n=1,2~,  \label{5b}
\end{equation}%
\begin{equation}
\underline{U}=%
\begin{pmatrix}
0 & x^{\left( 1\right) } & 0 & x^{\left( 2\right) } \\ 
\tilde{x}^{\left( 1\right) } & 0 & \tilde{y}^{\left( 1\right) } & 0 \\ 
0 & y^{\left( 1\right) } & 0 & y^{\left( 2\right) } \\ 
\tilde{x}^{\left( 2\right) } & 0 & \tilde{y}^{\left( 2\right) } & 0%
\end{pmatrix}%
~,  \label{5c}
\end{equation}%
as well as the following parameterization of the constant $4\otimes 4$
matrix $A$:%
\begin{equation}
\underline{A}=%
\begin{pmatrix}
\alpha & 0 & 0 & 0 \\ 
0 & a^{(11)}-\alpha & 0 & a^{(21)} \\ 
0 & 0 & \alpha & 0 \\ 
0 & a^{(12)} & 0 & a^{(22)}-\alpha%
\end{pmatrix}%
~.  \label{5d}
\end{equation}%
Note that we are considering a case with $N=2,~L=1.$

We then obtain the following \textit{integrable}, indeed \textit{solvable,}
system of \textit{four} coupled linear plus cubic oscillators in the plane: 
\end{subequations}
\begin{subequations}
\label{6}
\begin{equation}
\overset{..}{\overrightarrow{r}}^{(1)}+a^{(11)}\,\overrightarrow{r}^{\left(
1\right) }+a^{(21)}\,\overrightarrow{r}^{\left( 2\right) }=-\overrightarrow{r%
}^{\left( 1\right) }\,\left( \overrightarrow{\tilde{r}}^{\left( 1\right)
}\cdot \overrightarrow{r}^{\left( 1\right) }\right) -\overrightarrow{r}%
^{\left( 2\right) }\,\left( \overrightarrow{\tilde{r}}^{\left( 2\right)
}\cdot \overrightarrow{r}^{\left( 1\right) }\right) ~,  \label{6a}
\end{equation}%
\begin{equation}
\overset{..}{\overrightarrow{r}}^{(2)}+a^{(12)}\,\overrightarrow{r}^{\left(
1\right) }+a^{(22)}\,\overrightarrow{r}^{\left( 2\right) }=-\overrightarrow{r%
}^{\left( 1\right) }\,\left( \overrightarrow{\tilde{r}}^{\left( 1\right)
}\cdot \overrightarrow{r}^{\left( 2\right) }\right) -\overrightarrow{r}%
^{\left( 2\right) }\,\left( \overrightarrow{\tilde{r}}^{\left( 2\right)
}\cdot \overrightarrow{r}^{\left( 2\right) }\right) ~,  \label{6b}
\end{equation}%
\begin{equation}
\overset{..}{\overrightarrow{\tilde{r}}}^{(1)}+a^{(11)}\,\overrightarrow{%
\tilde{r}}^{\left( 1\right) }+a^{(12)}\,\overrightarrow{\tilde{r}}^{\left(
2\right) }=-\overrightarrow{\tilde{r}}^{\left( 1\right) }\,\left( 
\overrightarrow{\tilde{r}}^{\left( 1\right) }\cdot \overrightarrow{r}%
^{\left( 1\right) }\right) -\overrightarrow{\tilde{r}}^{\left( 2\right)
}\,\left( \overrightarrow{\tilde{r}}^{\left( 1\right) }\cdot \overrightarrow{%
r}^{\left( 2\right) }\right) ~,  \label{6c}
\end{equation}%
\begin{equation}
\overset{..}{\overrightarrow{\tilde{r}}}^{(2)}+a^{(21)}\,\overrightarrow{%
\tilde{r}}^{\left( 1\right) }+a^{(22)}\overrightarrow{\tilde{r}}^{\left(
2\right) }=-\overrightarrow{\tilde{r}}^{\left( 1\right) }\,\left( 
\overrightarrow{\tilde{r}}^{\left( 2\right) }\cdot \overrightarrow{r}%
^{\left( 1\right) }\right) -\overrightarrow{\tilde{r}}\,\left( 
\overrightarrow{\tilde{r}}^{\left( 2\right) }\cdot \overrightarrow{r}%
^{\left( 2\right) }\right) ^{\left( 2\right) }~.  \label{6d}
\end{equation}%
Further reductions to systems of less than \textit{four} $2$-vectors are
easy to obtain by appropriate reductions. For instance by setting to zero
the coefficient $a^{(21)}$ and the $2$-vector $\overrightarrow{\tilde{r}}%
^{\left( 2\right) }$ (and setting for notational convenience $a^{(11)}=a,$ $%
a^{(12)}=b,$ $a^{(22)}=c$ and $\overrightarrow{\tilde{r}}^{\left( 1\right) }=%
\overrightarrow{r}^{\left( 3\right) }$) we get the following system of 
\textit{three }coupled oscillators in the plane: 
\end{subequations}
\begin{subequations}
\label{7}
\begin{equation}
\overset{..}{\overrightarrow{r}}^{(1)}+a\,\overrightarrow{r}^{\left(
1\right) }=-\overrightarrow{r}^{\left( 1\right) }\,\left( \overrightarrow{r}%
^{\left( 1\right) }\cdot \overrightarrow{r}^{\left( 3\right) }\right) -%
\overrightarrow{r}^{\left( 2\right) }\left( \overrightarrow{\tilde{r}}%
^{\left( 2\right) }\cdot \overrightarrow{r}^{\left( 1\right) }\right)
~,\smallskip  \label{7a}
\end{equation}%
\begin{equation}
\overset{..}{\overrightarrow{r}}^{(2)}+b\,\overrightarrow{r}^{\left(
1\right) }+c\,\overrightarrow{r}^{\left( 2\right) }=-\overrightarrow{r}%
^{\left( 1\right) }\,\left( \overrightarrow{r}^{\left( 2\right) }\cdot 
\overrightarrow{r}^{\left( 3\right) }\right) ~,  \label{7b}
\end{equation}%
\begin{equation}
\overset{..}{\overrightarrow{r}}^{(3)}+a\,\overrightarrow{r}^{\left(
3\right) }=-\overrightarrow{r}^{\left( 3\right) }\,\left( \overrightarrow{r}%
^{\left( 3\right) }\cdot \overrightarrow{r}^{\left( 1\right) }\right) ~.
\label{7c}
\end{equation}%
By setting moreover to zero $\overrightarrow{r}^{\left( 2\right) }$ and $b,$
and setting $\overrightarrow{r}^{\left( 1\right) }=\overrightarrow{r},~%
\overrightarrow{r}^{\left( 3\right) }=\overrightarrow{\tilde{r}},$ we obtain
the following system of \textit{two} coupled oscillators in the plane: 
\end{subequations}
\begin{subequations}
\label{8}
\begin{equation}
\overset{..}{\overrightarrow{r}}+a\,\overrightarrow{r}=-\overrightarrow{r}%
\,\left( \overrightarrow{r}\cdot \overrightarrow{\tilde{r}}\right) ~,
\label{8a}
\end{equation}%
\begin{equation}
\overset{..}{\overrightarrow{\tilde{r}}}+a\,\overrightarrow{\tilde{r}}=-%
\overrightarrow{\tilde{r}}\,\left( \overrightarrow{r}\cdot \overrightarrow{%
\tilde{r}}\right) ~.  \label{8b}
\end{equation}%
And clearly the choice $\overrightarrow{\tilde{r}}=\overrightarrow{r}$
reduces this system to the (trivially solvable) single equation 
\end{subequations}
\begin{equation}
\overset{..}{\overrightarrow{r}}+a\,\overrightarrow{r}=-\overrightarrow{r}%
\,r^{2}~.  \label{9}
\end{equation}

A more general system of \textit{two} coupled oscillators than (\ref{8})
obtains from (\ref{6}) by setting $a^{(11)}=a,$ $a^{(12)}=a^{(21)}=b,$ $%
a^{(22)}=c,$ $\overrightarrow{\tilde{r}}^{\left( 1\right) }=\overrightarrow{r%
}^{\left( 1\right) },$ $\overrightarrow{\tilde{r}}^{\left( 2\right) }=%
\overrightarrow{r}^{\left( 2\right) }$: 
\begin{subequations}
\label{10}
\begin{equation}
\overset{..}{\overrightarrow{r}}^{(1)}+a\overrightarrow{r}^{\left( 1\right)
}+b\overrightarrow{r}^{\left( 2\right) }=-\overrightarrow{r}^{\left(
1\right) }\left( \overrightarrow{r}^{\left( 1\right) }\cdot \overrightarrow{r%
}^{\left( 1\right) }\right) -\overrightarrow{r}^{\left( 2\right) }\left( 
\overrightarrow{r}^{\left( 2\right) }\cdot \overrightarrow{r}^{\left(
1\right) }\right) ~,  \label{10a}
\end{equation}

\begin{equation}
\overset{..}{\overrightarrow{r}}^{(2)}+b\overrightarrow{r}^{\left( 1\right)
}+c\overrightarrow{r}^{\left( 2\right) }=-\overrightarrow{r}^{\left(
1\right) }\left( \overrightarrow{r}^{\left( 1\right) }\cdot \overrightarrow{r%
}^{\left( 2\right) }\right) -\overrightarrow{r}^{\left( 2\right) }\left( 
\overrightarrow{r}^{\left( 2\right) }\cdot \overrightarrow{r}^{\left(
2\right) }\right) ~.  \label{10b}
\end{equation}%
And of course from this system the single equation (\ref{9}) can be easily
obtained via an additional, obvious, reduction.

\bigskip

\subsection{Oscillators in ordinary ($3-$dimensional) space}

Let us consider the following parameterization of a $6\otimes 6$ matrix $%
\underline{U}$ (see (\ref{1})) in terms of \textit{six} $3-$vectors (see
(I.7)): 
\end{subequations}
\begin{subequations}
\label{11}
\begin{equation}
\underline{U}\doteq \left( \overrightarrow{r}^{\left( 1\right) },%
\overrightarrow{r}^{\left( 2\right) },\overrightarrow{r}^{\left( 3\right) };%
\overrightarrow{\tilde{r}}^{\left( 1\right) },\overrightarrow{\tilde{r}}%
^{\left( 2\right) },\overrightarrow{\tilde{r}}^{\left( 3\right) }\right) ~,
\label{11a}
\end{equation}%
\begin{equation}
\overrightarrow{r}^{\left( n\right) }\equiv \left( x^{\left( n\right)
},y^{\left( n\right) },z^{\left( n\right) }\right) ,~~~\overrightarrow{%
\tilde{r}}^{\left( n\right) }\equiv \left( \tilde{x}^{\left( n\right) },%
\tilde{y}^{\left( n\right) },\tilde{z}^{\left( n\right) }\right) ,~n=1,2,3~;
\label{11b}
\end{equation}%
\begin{equation}
\underline{U}=%
\begin{pmatrix}
0 & x^{\left( 1\right) } & 0 & x^{\left( 2\right) } & 0 & x^{\left( 3\right)
} \\ 
\tilde{x}^{\left( 1\right) } & 0 & \tilde{y}^{\left( 1\right) } & 0 & \tilde{%
z}^{\left( 1\right) } & 0 \\ 
0 & y^{\left( 1\right) } & 0 & y^{\left( 2\right) } & 0 & y^{\left( 3\right)
} \\ 
\tilde{x}^{\left( 2\right) } & 0 & \tilde{y}^{\left( 2\right) } & 0 & \tilde{%
z}^{\left( 2\right) } & 0 \\ 
0 & z^{\left( 1\right) } & 0 & z^{\left( 2\right) } & 0 & z^{\left( 3\right)
} \\ 
\tilde{x}^{\left( 3\right) } & 0 & \tilde{y}^{\left( 3\right) } & 0 & \tilde{%
z}^{\left( 3\right) } & 0%
\end{pmatrix}%
~,  \label{11c}
\end{equation}%
as well as the following parameterization of the constant $6\otimes 6$
matrix $A$:%
\begin{equation}
\underline{A}=%
\begin{pmatrix}
\alpha & 0 & 0 & 0 & 0 & 0 \\ 
0 & a^{(11)}-\alpha & 0 & a^{(12)} & 0 & a^{(13)} \\ 
0 & 0 & \alpha & 0 & 0 & 0 \\ 
0 & a^{(21)} & 0 & a^{(22)}-\alpha & 0 & a^{(23)} \\ 
0 & 0 & 0 & 0 & \alpha & 0 \\ 
0 & a^{(31)} & 0 & a^{(32)} & 0 & a^{(33)}-\alpha%
\end{pmatrix}%
~.  \label{11d}
\end{equation}%
Note that we are considering a case with $N=3,~L=1.$

We then obtain the following \textit{integrable}, indeed \textit{solvable},
system of \textit{six} coupled linear plus cubic oscillators in ordinary ($%
3- $dimensional) space: 
\end{subequations}
\begin{subequations}
\label{12}
\begin{equation}
\overset{..}{\overrightarrow{r}}^{(n)}=-\sum_{\nu =1}^{3}\left\{ 
\overrightarrow{r}^{(\nu )}\,\left[ a^{(\nu n)}+\left( \overrightarrow{%
\tilde{r}}^{(\nu )}\cdot \overrightarrow{r}^{(n)}\right) \right] \right\}
,~n=1,2,3\;,  \label{12a}
\end{equation}%
\smallskip 
\begin{equation}
\overset{..}{\overrightarrow{\tilde{r}}}^{(n)}=-\sum_{\nu =1}^{3}\left\{ 
\overrightarrow{\tilde{r}}^{(\nu )}\,\left[ a^{(n\nu )}+\left( 
\overrightarrow{\tilde{r}}^{(n)}\cdot \overrightarrow{r}^{(\nu )}\right) %
\right] \right\} ,~n=1,2,3\;.  \label{12b}
\end{equation}

It is plain that if $a^{(\nu n)}=a^{(n\nu )}$ one can set $\overrightarrow{%
\tilde{r}}^{(n)}=\overrightarrow{r}^{(n)},$ obtaining thereby the following
system of \textit{three} coupled linear plus cubic oscillators: 
\end{subequations}
\begin{equation}
\overset{..}{\overrightarrow{r}}^{(n)}=-\sum_{\nu =1}^{3}\left\{ 
\overrightarrow{r}^{(\nu )}\,\left[ a^{(\nu n)}+\left( \overrightarrow{r}%
^{(\nu )}\cdot \overrightarrow{r}^{(n)}\right) \right] \right\} ,~n=1,2,3\;.
\label{13}
\end{equation}

Reductions of the system (\ref{12}) to a number of oscillators smaller than 
\textit{six} can be simply obtained by setting to zero some\textit{\ }%
vectors and, for consistency, some of the constants $a^{(n\nu )}$. For
instance, by setting $a^{(31)}=a^{(13)}=a^{(32)}=a^{(23)}=0$ (and, for
notational convenience, $a^{(11)}=a,~a^{(21)}=b,~a^{(12)}=c,~a^{(22)}=d$)
and $\overrightarrow{r}^{(3)}=\overrightarrow{\tilde{r}}^{(3)}=0,$ one gets
the following system of \textit{four} coupled linear plus cubic oscillators
in ordinary ($3$-dimensional$)$ space:\medskip 
\begin{subequations}
\label{14}
\begin{equation}
\overset{..}{\overrightarrow{r}}^{(1)}+a\,\overrightarrow{r}^{(1)}+b\,%
\overrightarrow{r}^{(2)}=-\overrightarrow{r}^{(1)}\,\left( \overrightarrow{r}%
^{(1)}\cdot \overrightarrow{\tilde{r}}^{(1)}\right) -\overrightarrow{r}%
^{(2)}\,\left( \overrightarrow{r}^{(1)}\cdot \overrightarrow{\tilde{r}}%
^{(2)}\right) ~,  \label{14a}
\end{equation}%
\begin{equation}
\overset{..}{\overrightarrow{r}}^{(2)}+c\,\overrightarrow{r}^{(1)}+d\,%
\overrightarrow{r}^{(2)}=-\overrightarrow{r}^{(1)}\,\left( \overrightarrow{r}%
^{(2)}\cdot \overrightarrow{\tilde{r}}^{(1)}\right) -\overrightarrow{r}%
^{(2)}\,\left( \overrightarrow{r}^{(2)}\cdot \overrightarrow{\tilde{r}}%
^{(2)}\right) ~,  \label{14b}
\end{equation}%
\begin{equation}
\overset{..}{\overrightarrow{\tilde{r}}}^{(1)}+a\,\overrightarrow{\tilde{r}}%
^{(1)}+b\,\overrightarrow{\tilde{r}}^{(2)}=-\overrightarrow{\tilde{r}}%
^{(1)}\,\left( \overrightarrow{\tilde{r}}^{(1)}\cdot \overrightarrow{r}%
^{(1)}\right) -\overrightarrow{\tilde{r}}^{(2)}\,\left( \overrightarrow{%
\tilde{r}}^{(1)}\cdot \overrightarrow{r}^{(2)}\right) ~,  \label{14c}
\end{equation}%
\begin{equation}
\overset{..}{\overrightarrow{\tilde{r}}}^{(1)}+c\,\overrightarrow{\tilde{r}}%
^{(1)}+d\,\overrightarrow{\tilde{r}}^{(2)}=-\overrightarrow{\tilde{r}}%
^{(1)}\,\left( \overrightarrow{\tilde{r}}^{(2)}\cdot \overrightarrow{r}%
^{(1)}\right) -\overrightarrow{\tilde{r}}^{(1)}\,\left( \overrightarrow{%
\tilde{r}}^{(2)}\cdot \overrightarrow{r}^{(2)}\right) ~.  \label{14d}
\end{equation}

Note that, denoting by $\pi $ the plane identified by the requirement that
the two vectors $\overrightarrow{r}^{(1)}$ and $\overrightarrow{r}^{(2)}$
lie \textit{initially }(namely, at $t=0)$ in it, if the initial velocities $%
\overset{\cdot }{\overrightarrow{r}}^{(1)}(0)$ and $\overset{\cdot }{%
\overrightarrow{r}}^{(2)}(0)$ also lie in the same plane $\pi ,$ then the
two vectors $\overrightarrow{r}^{(1)}(t)$ and $\overrightarrow{r}^{(2)}(t)$
remain in this same plane $\pi $ throughout their time evolution --
independently of the behavior of the two vectors $\overrightarrow{\tilde{r}}%
^{(1)}$ and $\overrightarrow{\tilde{r}}^{(2)},$ which of course might as
well also always remain on some (possibly different) plane, say $\tilde{\pi}%
, $ if their initial velocities lie in the same plane defined by their
initial positions.

Further reductions to \textit{two}, or just \textit{one}, oscillators yield
equations (for $3$-vectors) analogous to those written above for $2$-vectors
(see (\ref{8}), (\ref{9}) and (\ref{10})).

\bigskip

\section{Hamiltonians}

The matrix evolution equation (\ref{1}) is Hamiltonian: it obtains, in the
standard manner, from the Hamiltonian 
\end{subequations}
\begin{equation}
H(\underline{P},\underline{U})=\text{trace}\left[ \frac{1}{2}\,\underline{P}%
^{2}+\underline{U}\,\underline{A}\,\underline{U}+\frac{1}{4}\,\underline{U}%
^{4}\right] ~,  \label{Ham}
\end{equation}%
where the canonical coordinates are the components of the matrix $\underline{%
U}$ and the corresponding canonical momenta are the components of the matrix 
$\underline{P}.$ (Indeed, the integrability of this matrix evolution
equation, (\ref{1}), is due to its being a reduction \cite{3} \cite{2} \cite%
{4} of the Non-Abelian Toda Lattice \cite{7} \cite{K}, the Hamiltonian
structure of which was already exploited in \cite{8}). This clearly entails
that all the Newtonian equations of motion reported in this paper (except (%
\ref{4})) are as well Hamiltonian.

We display here only the Hamiltonian function that, via the standard
Hamiltonian equations, yields the basic Newtonian equations of motion (\ref%
{2}). It reads:

\begin{equation*}
H\left( \overrightarrow{p}^{\left( n\ell \right) },\,\overrightarrow{\tilde{p%
}}^{\left( n\ell \right) };\overrightarrow{r}^{\left( n\ell \right) },\,%
\overrightarrow{\tilde{r}}^{\left( n\ell \right) }\right) \mathcal{=}%
\sum_{\nu =1}^{N}\sum_{\lambda =1}^{L}\left( \overrightarrow{p}^{\left( \nu
\lambda \right) }\cdot \overrightarrow{\tilde{p}}^{\left( \nu \lambda
\right) }\right)
\end{equation*}%
\begin{equation}
+\sum_{\nu ,\tilde{\nu}=1}^{N}\sum_{\lambda ,\tilde{\lambda}=1}^{L}\left[
a_{\lambda \tilde{\lambda}}^{(\nu \tilde{\nu})}~\overrightarrow{r}^{\left(
\nu \lambda \right) }\cdot \overrightarrow{\tilde{r}}^{\left( \tilde{\nu}%
\tilde{\lambda}\right) }+\frac{1}{2}\left( \overrightarrow{r}^{\left( \nu
\lambda \right) }\cdot \overrightarrow{\tilde{r}}^{\left( \tilde{\nu}\tilde{%
\lambda}\right) }\right) \left( \overrightarrow{r}^{\left( \tilde{\nu}\tilde{%
\lambda}\right) }\cdot \overrightarrow{\tilde{r}}^{\left( \nu \lambda
\right) }\right) \right] ~.  \label{ham}
\end{equation}%
In this Hamiltonian, the canonical coordinates are the $2\,N^{2}\,L$
components of the $2\,N\,L$ $\ N$-vectors $\overrightarrow{r}^{\left( n\ell
\right) },\overrightarrow{\tilde{r}}^{\left( n\ell \right) }$, $n=1,2,..,N$, 
$~\ell =1,2,..,L,$ and the corresponding canonical momenta are the $%
2\,N^{2}\,L$ components of the $2\,N\,L$ $\ N$-vectors $\overrightarrow{p}%
^{\left( n\ell \right) },\overrightarrow{\tilde{p}}^{\left( n\ell \right) }$%
, $n=1,2,..,N$, $~\ell =1,2,..,L.$ Note that this Hamiltonian entails that

\begin{equation}
\overset{.}{\overrightarrow{r}}^{(n\ell )}=\overrightarrow{\tilde{p}}%
^{\left( n\ell \right) }~,~~~\overset{.}{\overrightarrow{\tilde{r}}}^{(n\ell
)}=\overrightarrow{p}^{\left( n\ell \right) }~.  \label{17}
\end{equation}%
Also note that this Hamiltonian is of normal type, namely it features a
kinetic energy term depending quadratically only on the momenta and a
potential energy term depending (quadratically and quartically) only on the
coordinates.

\bigskip

\section{Final remarks}

As we already emphasized above, the Newtonian equations of motion considered
in this paper are all \textit{covariant} hence \textit{rotation-invariant},
this being a rather essential condition to interpret them as describing a
many-body problem. They are, however, not invariant under \textit{%
translations}; but variants of them that do possess this additional
property, without forsaking the property to be \textit{integrable} indeed 
\textit{solvable,} can be easily manufactured (for a technique to do so see 
\cite{4}). This generalization, as well as detailed explorations of the
actual behavior of the solutions of these models, are left as tasks for the
future; as well as the application of these findings to phenomenologically
interesting situations (in physics or elsewhere).

A more challenging task for the future -- hopefully appealing to physicists
and to mathematicians -- will be the study of the quantal counterparts of
the models discussed in this paper.

\end{document}